\documentclass[11]{article}
\usepackage[utf8]{inputenc}

\usepackage{amsmath}
\usepackage{amssymb}
\usepackage{graphicx}
\usepackage{color}
\usepackage{multicol}
\usepackage{caption,subcaption}
\usepackage{hyperref}
\usepackage{tikz}
\usetikzlibrary{quantikz}
\usepackage[breakable]{tcolorbox}

\numberwithin{equation}{section}

\newenvironment{system}[1][]%
	{\begin{eqnarray} #1 \left\{ \begin{array}{lll}}%
	{\end{array} \right. \end{eqnarray}}

\newcommand{\Frac}[2]{\displaystyle \frac{#1}{#2}}

\newcommand{\eps}{{\varepsilon}}
\newcommand{\boldTheta}{\boldsymbol{\theta}}

\usepackage{stackengine}
\def\deq{\mathrel{\ensurestackMath{\stackon[1pt]{=}{\scriptstyle\nabla}}}}

\usepackage{mathtools}

\begin{document}

\vspace{-0.3in}
\title{Quantum Natural Gradient with 
Efficient Backtracking Line Search
}
\vspace{-0.5in}

\author{Touheed Anwar Atif,  Uchenna Chukwu  \\
 Jesse Berwald and Raouf Dridi \\
 ~~\\
Quantum Computing Inc.}

\maketitle

\newcommand{\jesse}[1]{\textcolor{magenta}{Jesse: #1}}












\begin{abstract}
 We consider the Quantum Natural Gradient Descent (QNGD) scheme which was recently proposed to train variational quantum algorithms. QNGD is Steepest Gradient Descent (SGD) operating on the complex projective space equipped with the Fubini-Study metric.  Here we present an adaptive implementation of  QNGD based on Armijo's rule, which is an efficient backtracking line search that enjoys a proven convergence. 
The proposed algorithm is tested using noisy simulators on three different models with various initializations. Our results show that Adaptive QNGD dynamically adapts the step size and consistently outperforms the original QNGD, which requires knowledge of optimal step size to {perform competitively}. In addition, we show that the additional complexity involved in performing the line search in Adaptive QNGD is minimal, ensuring the gains provided by the proposed adaptive strategy dominates any increase in complexity. Additionally, our benchmarking demonstrates that a simple SGD algorithm (implemented in the Euclidean space) equipped with the adaptive scheme above, can yield performances similar to the QNGD scheme with optimal step size. 
 
  Our results are yet another confirmation of the importance of differential geometry in variational quantum computations.  As a matter of fact, we foresee advanced mathematics to play a prominent role in the NISQ era in guiding the design of faster and more efficient algorithms.

\end{abstract}

\section{Introduction}

 Quantum processors currently suffer from low qubit counts and short {coherence times}, which restricts the number (and the depth) of {quantum} algorithms that can be executed. An approach to deal with this limitation is the  {\it variational approach} \cite{farhi2014quantum, Peruzzo_2014, mcclean2016theory}, where the short-lived qubits are {\it recycled} multiple times. A classical optimizer  (e.g., SGD) is entrusted with administering part of the dynamics. The quantum component runs a parametric circuit, an {\it ansatz}, given the parameter values provided by the classical controller.  Conversely, the  classical optimizer, updates the circuit parameters by minimizing the expectation value of the parametric circuit. The term hybrid quantum-classical neural network is often used and puts variational algorithms in the wider context of quantum machine learning \cite{McClean_2016, broughton2020tensorflow}.

A recent and attractive proposal to train variational algorithms was outlined in \cite{Stokes_2020}, which employs a SGD, preconditioned with the inverse of the  Fubini-Study tensor. For a geometer,  the parameters update takes place on the K\"ahler manifold of the pure quantum states (i.e., the complex projective space).  The idea was used before in \cite{amari} in classical machine learning. Subsequent results (\cite{Stokes_2020,yamamoto2019natural, Wierichs_2020}) showed that this preconditioning does indeed lead to a better performance and faster training.

Here we take this geometrical approach further and provide
an adaptive implementation of the QNGD scheme, thus mitigating the dependence of its performance  on the step size. For this, we have employed the so-called Armijo rule, which is an efficient backtracking line search scheme with a proved convergence \cite{doi:10.1137/040605266}. 
We have tested our scheme on three different models with various initializations. Our results show that the Adaptive QNGD scheme dynamically adapts the step size and consistently outperforms QNGD which requires the knowledge of the optimal step size to perform competitively.  Remarkably, our results also show that 
 a simple steepest gradient descent  equipped with the adaptive scheme above, can yield performances similar to the QNGD with optimal step size.

This work is organized as follows. In Section \ref{sec:QNGD}, we briefly review the concepts behind the QNGD technique, and  highlight 
its implementation on quantum processors. In  Section \ref{sec:adaptQNGD}, we introduce our scheme  and  compare its complexity with that of the QNGD scheme. We show that the additional circuit evaluations needed  are insignificant.
Section \ref{sec:simResults} contains the results of our benchmarking conducted on three different problems. We conclude with a brief summary.
 
  \tableofcontents

\newpage 
\section{Acronyms}
\begin{center}
\begin{tabular}{ |l|l| } 
 \hline
  {\sc AdaptQNGD } &  Adaptive Quantum Natural Gradient Descent \\ 
 \hline
 {\sc qngd} &   Quantum Natural Gradient Descent \\ 
 \hline
 {\sc sgd} &   Steepest
Gradient Descent \\ 
 \hline
 {\sc realAmplitude} & ansatz alternating $R_Y(\theta)$ and {\sc cnot} layers\\
 \hline
  {\sc FullFubiniMetric} & full Fubini-Study metric calculation procedure\\
 \hline
\end{tabular}
\end{center}

\section{Quantum Natural Gradient Descent}\label{sec:QNGD}
    
The  complex projective space
$\mathbb C\mathbb P^N$ is identified with the quotient space $S^{2N+1}/U(1)$ where $U(1)$ is the circle group. 
Given two (normalized) pure quantum states $|\psi\rangle$ and $|\varphi\rangle$ in $\mathbb C\mathbb P^N$,  the Fubini-Study distance between  the two is given by
\begin{equation}
    d_{FS} (|\psi\rangle, |\varphi\rangle) 
    = \mathrm{arccos} \sqrt{ {\langle \psi| \varphi\rangle \langle\varphi |\psi\rangle}{}}.
\end{equation}
Let $N=2^n$. For convenience,  we place   ourselves in $\mathbb C\mathbb P^{N-1}$. Let us assume    $|\varphi\rangle = U(\boldTheta) |0\rangle,$ { where the {\it ansatz }}  $ \{U(\boldTheta), \, \boldTheta\in  \mathbb R^p\}$ is a smooth submanifold of the Lie group $U(2^n)$.  
Passing to  the infinitesimals gives
 the {\it quantum geometric tensor}  \cite{Stokes_2020}
\begin{equation}\label{FS}
    G_{ij} = \left\langle\partial_{\boldTheta_i}\varphi | \partial_{\boldTheta_j}\varphi \right\rangle
    - \left\langle \partial_{\boldTheta_i}\varphi | \varphi \right\rangle
     \left\langle \varphi| \partial_{\boldTheta_j}\varphi \right\rangle. 
\end{equation}
The real part $F=\mathcal{R}e(G)$, called the {\it Fubini-Study metric}, is positive semi-definite, Hermitian, and is the unique metric compatible with the quotient structure above. Formally, $\mathbb C\mathbb P^N$ equipped with the Fubini-Study metric is a K\"ahler manifold.  This rich geometrical structure implies that the gradient descent update has the form 
\begin{equation}
    \boldTheta  \leftarrow \boldTheta - \lambda \nabla f(\boldTheta),
\end{equation}
with
\begin{equation}
    \nabla f(\boldTheta) =  F(\boldTheta)^{-1}\nabla_{\mathrm{Eucl}} f(\boldTheta),
\end{equation}
for any smooth real-valued function $f$ on $\mathbb C\mathbb P^{N-1}$. 


All ansatzes considered in this paper are 
structured with layers. An example of this is given by the following parametric circuit
\begin{center}
\begin{quantikz}
\lstick{$\ket{0}$} & \gate{R_Y(2a)} & \ctrl{1} & \gate{R_Y(2c)} \qw \\
\lstick{$\ket{0}$} & \gate{R_Y(2b)} & \targ{} & \gate{R_Y(2d)}\qw \\
\end{quantikz}  
\end{center}
which has two parametric layers: a first layer of {parametric} rotations, $R_Y(\theta) = e^{-i\frac{\theta}{2}Y}$, followed by a non-parametric layer consisting of a single {\sc cnot} gate, { followed by a final parametric} layer of $Y$ rotations. This type of ansatze (alternating is $R_Y(\theta)$ {\sc cnot} gates) is referred to as a {\sc realAmplitude} ansatz in the literature. 

Given this layer-based structure, the Fubini-Study metric is structured into block-diagonal (a block for each layer)  and off-diagonal parts. The former are readily computed as in  \cite{Stokes_2020}. For the off-diagonal parts, which are slightly less straightforward to implement, we have used
a method similar to the one proposed in \cite{romero2018strategies} which adds an extra ancilla qubit to the original circuit. For instance, continuing with the ansatz above,  the  circuit for the first component of the off-diagonal term   $G_{ac} = \left\langle\partial_a\varphi | \partial_c\varphi \right\rangle$ is given by 

\begin{center}
\begin{quantikz}
\lstick{$\ket{0}$} & \qw  & \gate{Y}    & \gate{R_Y(2a)} & \ctrl{1}& \gate{Y}& \qw \qw \\
\lstick{$\ket{0}$} & \qw  & \qw         & \gate{R_Y(2b)} & \targ{} & \qw & \qw \qw \\
\lstick{$\ket{0}$} &\gate{H} &\ctrl{-2} & \qw            &\qw      &\ctrl{-2}&\gate{H}\qw
\end{quantikz}  
\end{center}
{Here, the second component in equation \ref{FS} vanishes, since the prepared quantum state has only real amplitudes.}
We shall refer  to the full Fubini-Study metric calculation procedure as {\sc FullFubiniMetric}.    

\section{Adaptive Quantum Natural Gradient Descent} \label{sec:adaptQNGD}

\subsection{Illustrative Example}\label{h2example}

Before we give the details of the proposed scheme, let us illustrate how the new adaptive scheme compares with traditional  QNGD scheme.
We will do so using the  problem of
finding the electronic ground state energy  (after Born-Oppenheimer approximation) of  the
Hydrogen molecule \cite{aszabo82:qchem}.  Its fermionic Hamiltonian is mapped (using Jordan–Wigner transformation) into the two-qubit Hamiltonian 
\begin{align} \label{eq:h2hamiltonian}
    H = \alpha_0 I_1 I_2 + \alpha_1 Z_1I_2 + \alpha_2 I_1Z_2 + \alpha_3 Z_1Z_2 + \alpha_4 Y_1Y_2 + \alpha_5 X_1X_2.
\end{align}
$I_i$,  $X_i, Y_i$ and $Z_i$ are resp., the identity operator and the Pauli operators acting on the $i$th qubits. The coefficients $\alpha_j$ are real-valued functions of the inter-atomic distance between the two Hydrogen nuclei. 
For this Hamiltonian, we consider the simplified ansatz 
\begin{center}
\begin{quantikz}
\lstick{$\ket{0}$} & \gate{R_Y(\theta_1)} & \ctrl{1} & \qw \qw \\
\lstick{$\ket{0}$} &
 \gate{R_Y(\theta_2)} & \targ{} &  \qw \qw \\
\end{quantikz}  
\end{center}
which prepares the state $|{\psi(\theta_1,\theta_2)}\rangle$. The Fubini-Study metric is



    \begin{align*}
    F(\theta_1,\theta_2) = \frac{1}{4}\left[\begin{array}{cc}
1 & 0 
\\
 0 & 1
\end{array}\right],
    \end{align*}
which is, in this simple case, independent of both $\theta_1$ and $\theta_2$.  

\begin{figure}[hbt!]
    \centering
        \includegraphics[scale=0.49]{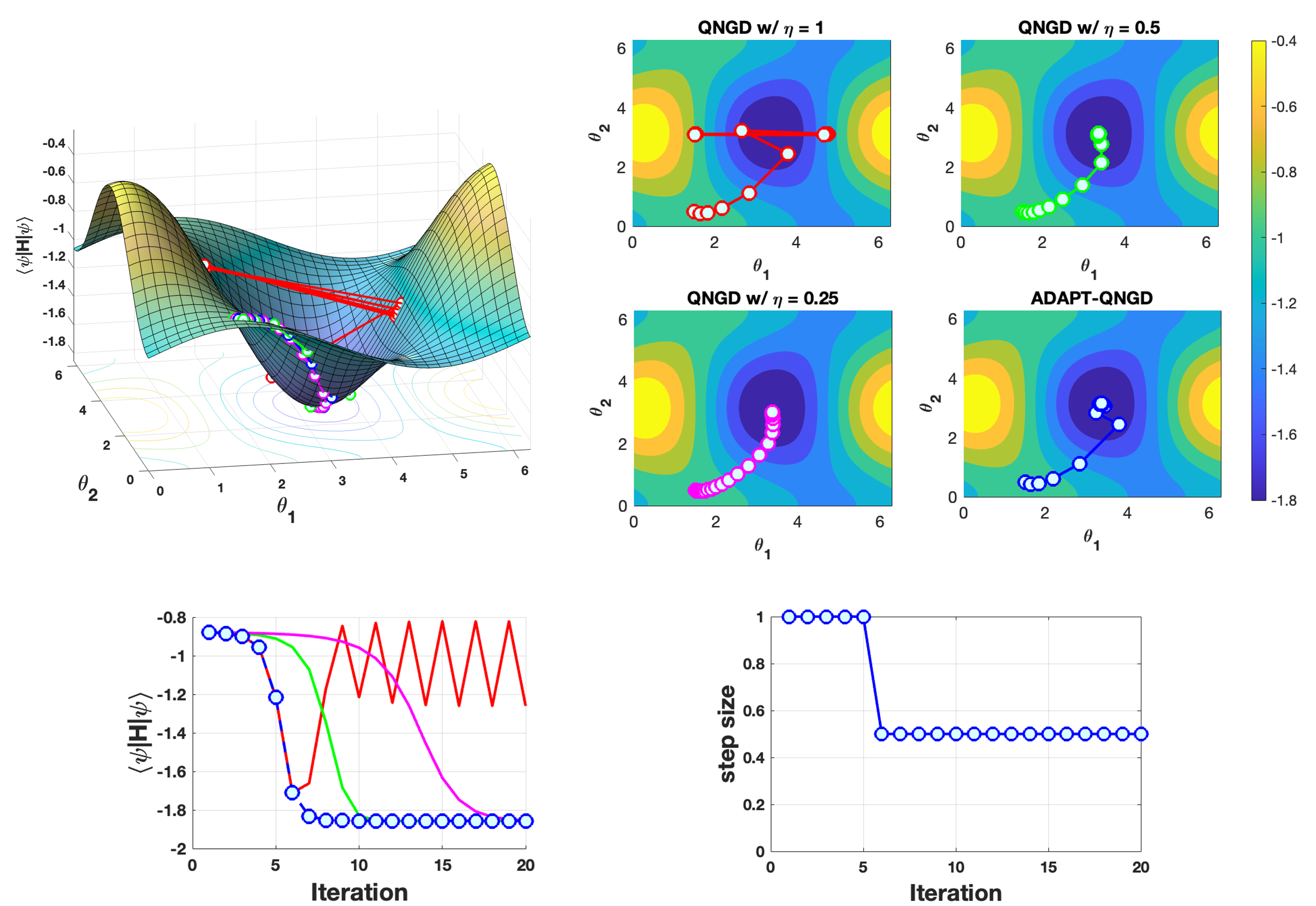}
    \caption{
    {\small 
    {\bf Upper Left}: The contour plot of the Hydrogen Hamiltonian with demo ansatz.  {\bf Upper Right}: Comparison of training paths taken by different schemes to approach the minima, with maximum number of epochs set to 20. The QNGD scheme with step size $\eta =1 $ does not converge. The QNGD with step sizes $=0.5$ and $0.25$ and the {\sc AdaptQNGD} schemes all converge, however, {\sc AdaptQNGD}  converges much faster. {\bf Lower left}: The figure shows the convergence of average value of the Hamiltonian with iteration counts for each of the schemes discussed in the center plot. {\bf Lower right}: The figure depicts the step sizes chosen by the {\sc AdaptQNGD} over the course of the algorithm. These figures demonstrate that the algorithm is able to exploit larger step sizes when the surface allows, and tune it lower upon reaching close to the minima, resulting in faster overall convergence. 
    }
    }
    \label{fig:H2_contour} 
\end{figure}

Figure \ref{fig:H2_contour} shows the training trajectories taken by QNGD with different learning rates $\eta$ and by the proposed adaptive scheme, all starting from the same initial point   $\theta_1 =-0.1$ and $\theta_2 = -0.2$.   The results show that, 
not only does {\sc AdaptQNGD}  converge much faster than the three other schemes, but it also uses a large step-size in the early stages of the algorithm (when the surface allows for this behaviour) and decreases the steps size closer to the minima, effectively slowing down which avoids the arbitrary bouncing as observed with step size $1$.


\newpage

\noindent \subsection{Main Algorithm}

\begin{itemize}  
    \item[] {\bf {\sc AdaptQNGD } algorithm}
    \item[] {\bf Input parameters}: $(\alpha,\beta,k_m,tol,\epsilon)$ -- See table below for description. 
    \item[]{\bf Step 1} Initialize: $\boldTheta_0, i=0.$ 
    \item[]{\bf Step 2} Compute the Euclidean gradient:
        \begin{itemize}
            \item [] $\nabla_{\mathrm{Eucl}} f(\boldTheta_i) = $ {\sc ParameterShift}$(\boldTheta_i)$.
        \end{itemize}
                
    \item[]{\bf Step 3}  Compute the Fubini-Study gradient: 
        \begin{itemize}
            \item[] $F(\boldTheta_i)$ =  {\sc FullFubiniMetric}$(\boldTheta_i)$,
            \item[] $F(\boldTheta_i)^{-1}$ = {\sc pseudoInvert}  $\left(F(\boldTheta_i), \epsilon\right)$, 
            \item[] $ \nabla f(\boldTheta_i) :=  F(\boldTheta_i)^{-1}\nabla_{\mathrm{Eucl}} f(\boldTheta_i)$.
        \end{itemize}
     
    \item[]{\bf Step 4} Terminate if the stopping criteria (defined below) is met, else
    \begin{itemize}
        \item [(1)]  Compute $k_i$:
\begin{align} 
    k_i &\deq \min\left\{k\in[0,k_{m}]\big|
    f(\boldTheta_i) - f(\boldTheta_i - \frac{\beta}{2^k}\nabla f(\boldTheta_i)) \geq \alpha\frac{\beta}{2^{k}}\|\nabla f(\boldTheta_i)\|_2^2) \right\}, \label{eq:armijos}
\end{align}
where $\|\|_2$ is the Euclidean norm.
    \item[(2)] Update step size to $\lambda_i = \beta/2^{k_i}$,
    \item[(3)] Update parameters as $\boldTheta_{i+1} = \boldTheta_i - \lambda_i \nabla f(\boldTheta_i),$ and proceed to Step 2.
    \end{itemize}

\end{itemize}
The five input parameters are defined as follows:
\begin{enumerate}
    \item $\alpha \in (0,1) $ is a constant to tune the sensitivity of the Armijo's line search principle. Lower values of $\alpha$ skews the rule toward preferring higher step sizes and higher skews it toward lower step sizes.
    \item $\beta > 0$ is the maximum rate used by the algorithm.
    \item $k_{m} > 0$ is the maximum number of steps searched by the line search algorithm in finding the best step size.
    \item  $tol > 0$  governs the stopping criteria defined below.
    \item $\epsilon > 0$ a tolerance parameter used in the {\sc pseudoInvert} function described below. For all our simulations, $\epsilon$ is set to $10^{-3}$.
\end{enumerate}

The {\sc ParameterShift} procedure in Step 2 is well known and defined in different places, including~\cite{Schuld_2019}.  The {\sc pseudoInvert} function is a variant of the Moore–Penrose pseudo-inverse  where apart from the zero eigenspace, we also ignore the eigenspace having eigenvalues less than $\epsilon$. 
It doesn't  satisfy all Moore–Penrose properties\cite{campbell2009generalized} but works very well in the simulations considered here, avoiding the unreasonably large jumps caused at the singularities of the Fubini metric $F$. This can also be seen as a variant of Tikhonov regularization \cite{golub1999tikhonov}, which has shown to circumvent the effect of barren plateaus \cite{Wierichs_2020}.      

\medskip

The stopping criteria is defined by the difference between the computed  and the exact energies i.e., $|f(\boldTheta)-f_{\mathrm{Exact}}(\boldTheta)|\leq tol$. In cases when computing the exact energies is infeasible, one can resort to other techniques developed in the gradient descent literature, for instance  
$||\nabla f(\boldTheta_i)|| \leq tol$, or even $\|f(\boldTheta_{i-1}) - f(\boldTheta_i))\| \leq tol$.

\medskip

The key point to note here is that step 4 shows that the line search is essentially a binary search. This means that our  scheme {\sc AdaptQNGD} compares  reasonably with the ``vanilla" {\sc QNGD} which requires  
$O(pM+p^2M) = O(p^2M)$ QPU calls--where $M$ is the total number of iterations to achieve the desired tolerance, and $p$ is  the total number of parameters.  For {\sc AdaptQNGD} the total number of QPU calls is $O(p^2M'+M'\log_2(k_m))$ or  $O(p^2M')$ since
$\log_2(k_m) < p^2$, where $M'$ (which is less or equal to $M$) 
is the number of iterations the proposed scheme takes to converge. This favourable asymptotic behaviour is key for the future applicability of our adaptive scheme. 

\section{Experimentation} \label{sec:probSetup}


\subsection{Setup}\label{sec:simDetails}
We benchamarked our algorithm on three models, the Hydrogen and Lithium Hydride molecules and the transverse field Ising model. We have  used the {\sc RealAmplitude} ansatze introduced in Section \ref{sec:QNGD}. All procedures used were implemented with TensorFlow Quantum library.
 
\subsubsection{The  Hydrogen Model} We took a closer look into the Hydrogen molecule discussed in the illustrative example in Subsection \ref{h2example}.  In particular, we wanted to compare against the results presented in~\cite{yamamoto2019natural}. Therefore,  we have used the same Hamiltonian (with $\alpha_0 = \alpha_3 = \alpha_4 = 0, \alpha_1  = \alpha_2 = 0.4, $ and $\alpha_5 = 0.2$)
\begin{equation}
    H = 0.4 (ZI + IZ) + 0.2 XX, 
\end{equation}
which has the spectrum $\{-\sqrt{17}/5, -1/5, 1/5, \sqrt{17}/5 \}$, with  $-\sqrt{17}/5 \approx -0.82462$ being the lowest energy (our target).  We have also considered the same ansatz 

\begin{center}
\begin{quantikz}
\lstick{$\ket{0}$} & \gate{R_Y(2a)} & \ctrl{1} & \gate{R_Y(2c)} \qw \\
\lstick{$\ket{0}$} & \gate{R_Y(2b)} & \targ{} & \gate{R_Y(2d)}\qw \\
\end{quantikz}  
\end{center}

We randomly initialize its parameters by 
    choosing 100 different initial points, uniformly distributed over $[0,2\pi]$. For each of these initial points, we first simulate the regular QNGD scheme for different rates, and then simulate the {\sc AdaptQNGD} scheme with $\alpha = 0.01$ and ${\beta = 0.5}$.


\subsubsection{Lithium Hydride (LiH) Model}
The Hamiltonian for LiH operates on four qubits and has 99 4-local Pauli terms. The complete list of these local Pauli  terms can be found in \cite{Kandala_2017}. 
We employ again the {\sc RealAmplitude} ansatz, now   with six parameterized layers, with a full entangling layer between every two of the former. The number of layers, compared to the H$_2$ model, is greater here to make the ansatz sufficiently expressive \cite{Kandala_2017}.   To initialize the parameters of the ansatz, a similar random strategy as above is used, with simulations for 100 different initial points. For each of these initial points, the QNGD scheme is simulated with different step sizes, and finally the {\sc AdaptQNGD}   is simulated using $\alpha = 0.01$ and $\beta = 1$.

~ 
\subsubsection{Transverse Field Ising (TFI) Model}
The Hamiltonian is used in \cite{Wierichs_2020} and  is given by 
\begin{equation}
    H_{TFIM} = - \left(\sum_{1\leq i\leq N} Z_i Z_{i+1} + t \sum_{1\leq i\leq N} X_i
    \right)
\end{equation}
with $t$, a real positive parameter, and the periodic condition, $N+1=1$, on the two sums (the $N$ here is a different notation than the one used in the Section \ref{sec:QNGD}). 

\medskip

    We fix $t$ to $0.1$ and  vary $N$ in the interval $[2,10]$. We have used two setups:  
    \begin{itemize}
        \item  In the first setup, we choose the ansatz to have a minimum number of parameters while being expressive enough. Our simulations show that we need at least two parameterized layers with a full entanglement layer in-between.
        \item  For the second setup, we choose the ansatz to study the behaviour of the proposed scheme in handling the over-parameterization problem. For this, we choose the number of parameterized layers equivalent to the size of the problem $(N)$, while enclosing an entanlging layer between each parameterized layers.  For instance, for $N=5$, we use five parameterized layers and four entangling layers.
    \end{itemize}

\subsection{Results} \label{sec:simResults}
In this section, we report our results and findings obtained using {\sc AdaptQNGD}   for the different models introduced above. Below we have set $\alpha = 0.01, \beta = 0.5$ and $k_{m} = 6$, independent of the model, the instance size and the ansatz depth.  The common thread below is that slight deviations in the optimal step size substantially degrades the performance of the various QNGD algorithms. At the same time, our {\sc AdaptQNGD} scheme was able to automatically adjust its step size and match the optimal QNGD. Moreover, the {\sc AdaptQNGD} was able to capture the correct dynamical change of the step size throughout the training and thus outperforming QNGD even with the optimal initial step size.

\subsubsection{The  Hydrogen Model}
\label{sec:h2contourplots}

\begin{figure}
    \centering
        \includegraphics[scale=0.47]{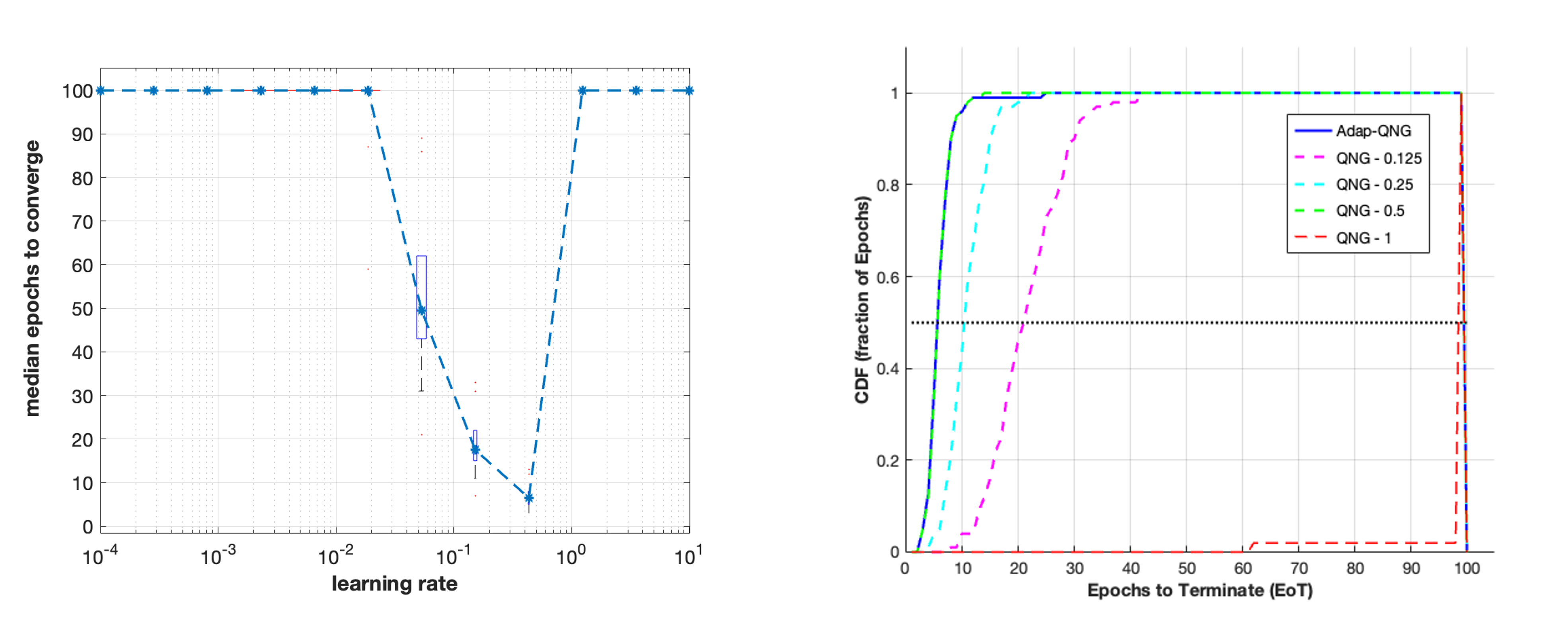}
    \caption{Left: Variation of the median number of epochs needed to converge for 10 random initializations of the {{H}$_2$} ansatz parameters with learning rate. The box plots indicates the 25th and the 75th percentile of the epochs required to converge. Right: Comparison of CDFs of the number of epochs needed for the ADAPT QNGD  and the QNGD, with varying step-sizes, to reach the lowest energy of the Hydrogen Hamiltonian.}
    \label{fig:H2_CDF}
\end{figure}

sAs discussed earlier, we chose $100$ different initial points. For each of these initial points, we simulate the QNGDS for step sizes $0.125, 0.25, 0.5$ and $1$.
We set the termination tolerance $tol$ to  $0.01$. We refer to  the number of epochs, for a given scheme to terminate, as Epochs to Terminate (EoT). 

Figure \ref{fig:H2_CDF} (Left) shows the variation of the median number of epochs needed to converge to the minimum eigenvalue of the Hydrogen Hamiltonian with the learning rate (step size) of the QNGD scheme. The median is taken across 10 random initial runs. It is clear from the plot that only for a range of learning rates, the QNGD scheme actually converges, and the rate of convergence highly depends on the chosen learning rate.

Figure \ref{fig:H2_CDF} (Right) shows the Cumulative Distribution Functions (CDF) of EoT for the Hydrogen model using the QNGDs with different step sizes and the {\sc AdaptQNGD}. In addition, the dotted black line at ordinate $0.5$ is drawn to compare the performance of different schemes at their medians. Figure~\ref{fig:H2_CDF}  clearly shows the dependence of the performance of the QNGD scheme on the choice of the step-size, with the step size of $0.5$ performing the best among others simulated. Doubling the step size to $1$ degrades the performance extensively causing almost all the runs to be requiring more than $100$ EoTs. On the other hand, decreasing the step sizes to $0.25$ and $0.125$  leads to sub-optimal performance. In general, 
it  is  difficult to predict beforehand the optimal step size, which requires extensive numerical simulations. 

Figure \ref{fig:H2_CDF} also shows  that the {\sc AdaptQNGD}, with the parameters mentioned above, performs similar to the QNGDs with the optimal step size $0.5$. Note that the {\sc AdaptQNGD}   only requires the step sizes it has to choose from, eliminating the need for the knowledge of the optimal step size. 


\subsubsection{Lithium Hydride Model}

\begin{figure}
\begin{subfigure}[b]{0.49\textwidth}
	\begin{centering}
	\includegraphics[width= \textwidth]{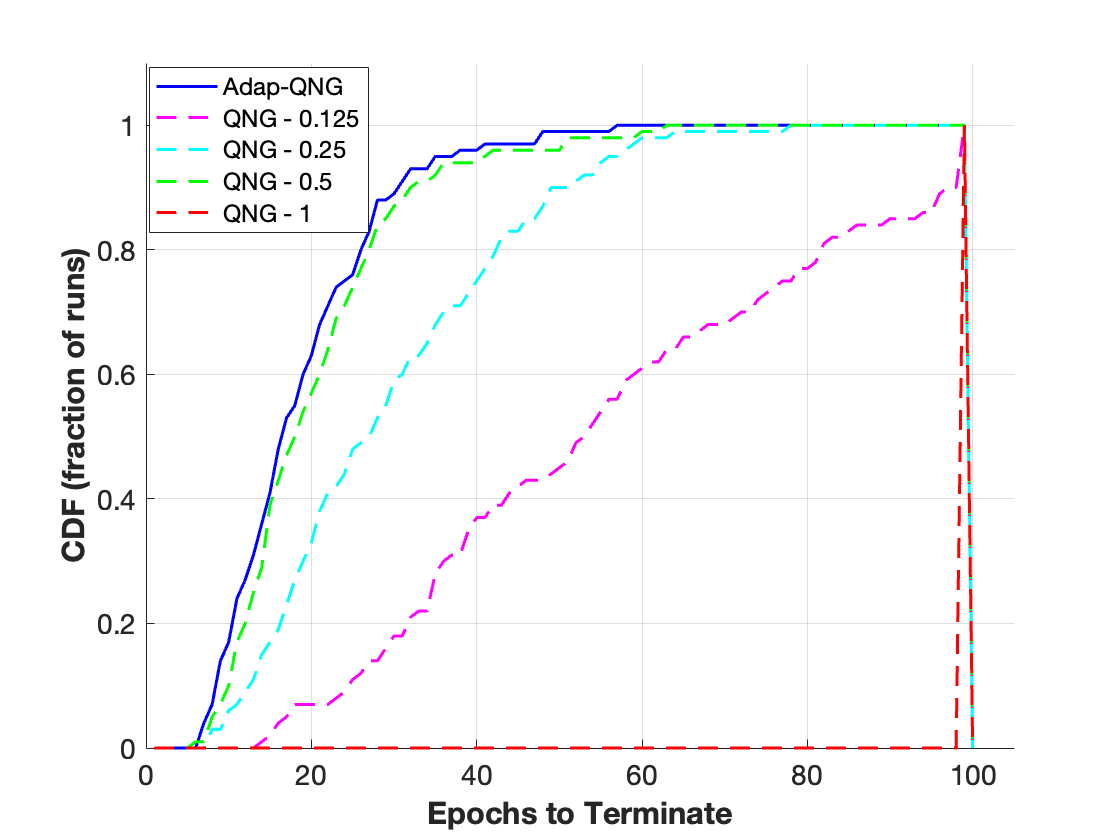}
	\par
	\end{centering}
    \caption{}
    \label{fig:LiHResult1}
 \end{subfigure}
 \begin{subfigure}[b]{0.49\textwidth}
	\begin{centering}
	\includegraphics[width= \textwidth]{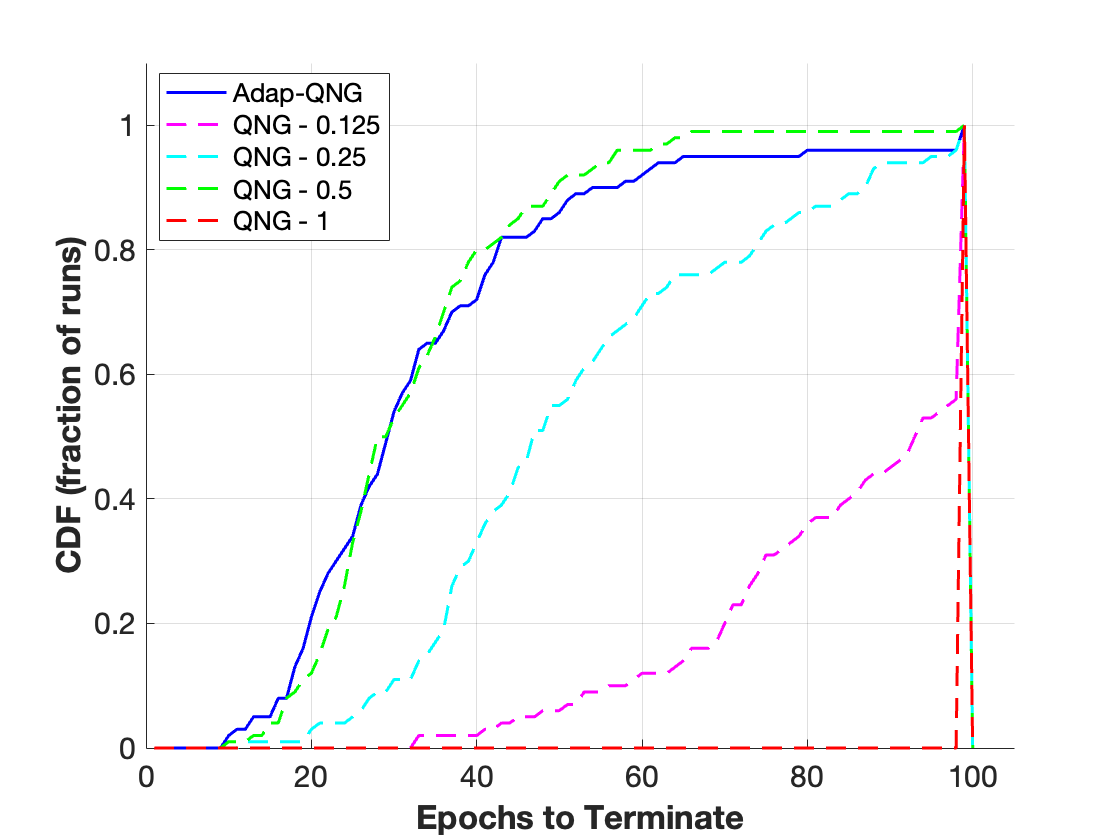}
	\par
	\end{centering}
    \caption{}
    \label{fig:LiHResult2}
 \end{subfigure}
	\caption{Comparison of CDFs of the number of epochs needed for the {\sc AdaptQNGD}  and the QNGDS to reach the lowest energy of the LiH Hamiltonian, with (a) $tol = 0.01$ and (b) $tol = 0.001$.}
\end{figure}
 
The simulations were ran for $100$ different random initial points, as discussed earlier, and for each initial point, the QNGDS was simulated with step sizes $0.125, 0.25, 0.5$ and $1$.
Figures \ref{fig:LiHResult1} and \ref{fig:LiHResult2} shows the distribution of EoT for each of the schemes, using $tol = 0.01$ and $tol = 0.001$, respectively.

Both figures   confirm the dependence of the performance of  the  QNGD scheme on the choice of the step-size. Notice that in this model, the optimal step size is $0.5$.
In the same time, 
 the {\sc AdaptQNGD}  was able to perform close to the best performance obtained by the QNGD scheme.
 In fact, the performance in Figure \ref{fig:LiHResult1}, of the former is slightly better than the latter. This is due to the fact that, although the step size of $0.5$ seems to be a better choice among the others simulated here, it may be the case that it is not the best for the entire duration of convergence. The landscape of the cost function may at times allow for a faster descent (or convergence), but the QNGD scheme is always limited to using only a fixed step size for the entire duration of the algorithm. Here again, the {\sc AdaptQNGD} learns this dynamically and proceeds with choosing larger step sizes than $0.5$ whenever possible, and as a result can converge faster.

\subsubsection{Transverse Field Ising (TFI) Model}
\begin{figure}[h]
    \centering
    \includegraphics[scale=0.2]{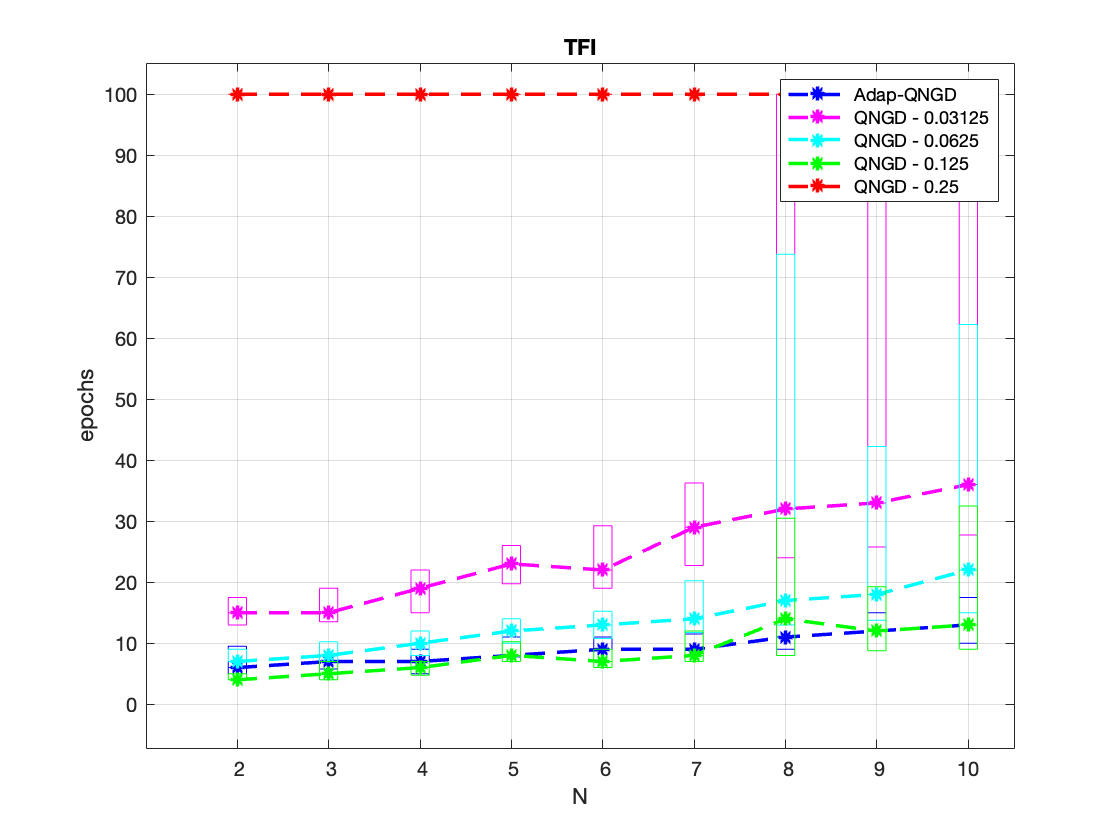}
    \caption{Comparison of median number of epochs needed of the {\sc AdaptQNGD}  with the QNGD scheme   using Boxplots. The box around the median points indicates the containment of 25th to 75th percentiles .}
    \label{fig:TFI_median}
\end{figure} 

In this last model, the simulations were ran for $25$ different random initial points, and for each initial point the QNGD scheme was simulated with step sizes $0.03125, 0.0625, 0.125,$ and $0.25$ and, as in the previous models, the {\sc AdaptQNGD} scheme was simulated with $\alpha= 0.01, \beta = 1$ and $k_{m} = 6.$  

For each size $N$, we first plot the median of the distribution of epochs to terminate, and then around the median, plot a box containing the realizations between the $25$th and the $75$th percentiles to obtain a comprehensive view of the performance.
Figure \ref{fig:TFI_median} shows our first result for the TFI model obtained using~${tol = 0.01}$.

This set of simulations again show that QNGD scheme depends on the choice of step size, in which the optimal step size in this case equals
 $0.125$. The performance degrades  considerably  
 with step size $0.25$ or higher. Expectedly, the {\sc AdaptQNGD}   overcomes this and adaptively chooses the best step size.  Note that as the problem size increases, the number of epochs needed to terminate also increases, a phenomena also seen in \cite{Wierichs_2020}. 

\begin{figure}[h]
    \centering
    \includegraphics[scale=0.2]{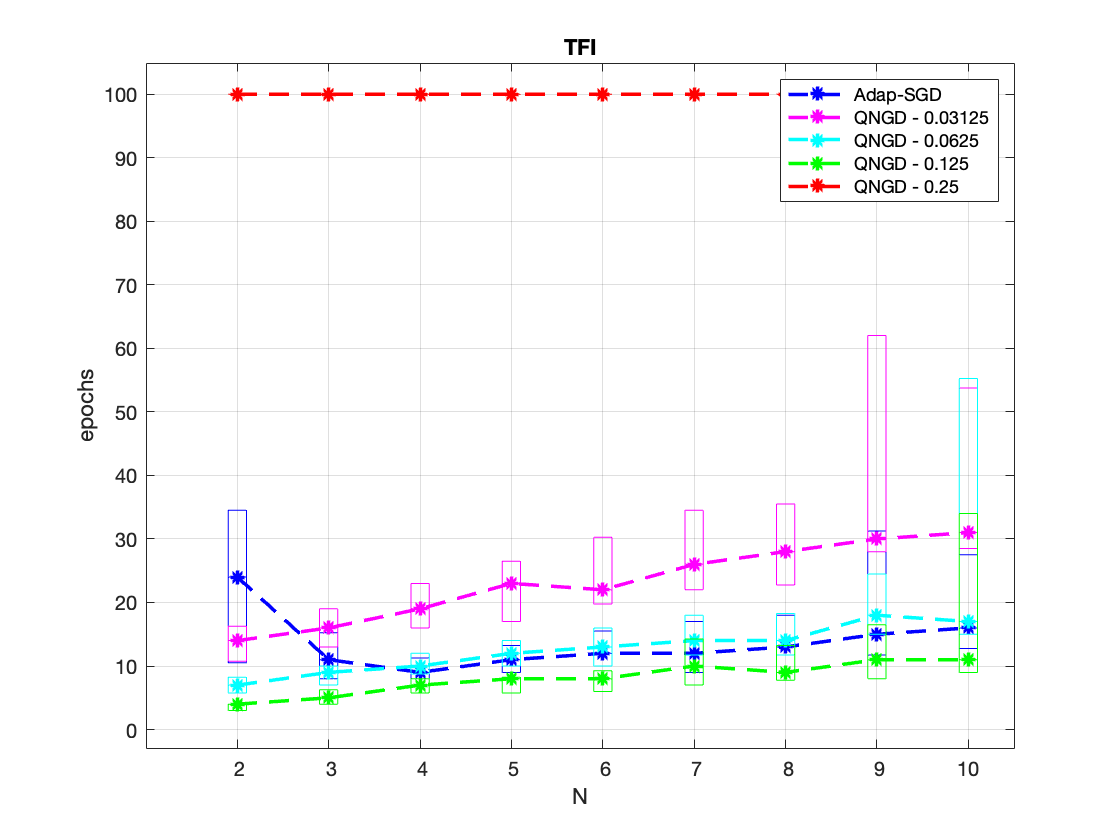}
    \caption{Comparison of median number of epochs needed of the ADAPT SGD scheme with the QNGD scheme  using Boxplots. The box around the median points indicates the containment of 25th to 75th percentiles .}
    \label{fid:TFI_SGD}
\end{figure} 

As a next result, we provide a comparison between the QNGD scheme and the Steepest Gradient Descent (SGD) equipped with the adaptive procedure used in {\sc AdaptQNGD} (i.e., setting the tensor $F$ to identity in {\sc AdaptQNGD}). This is an interesting comparison since the vanilla SGD  is blind to the underlying geometry (the complex projective space with the Fubini-Study metric), which results in
much slower convergence than the QNGD scheme \cite{yamamoto2019natural}. Remarkably, by using the adaptive step size technique, which defaults to choosing a large step size when permissible by the landscape, we are able to obtain similar performance as the optimal QNGD scheme. Figure \ref{fid:TFI_SGD} demonstrates this behaviour. This can prove to be very useful, as avoiding the computation of Fubini matrix leads to a substantial decrease in the number of circuit evaluations and henceforth the cost of simulating the algorithm can be significantly smaller.


\subsubsection{Over-parameterization}

\begin{figure}[h]
    \centering
    \includegraphics[scale=0.2]{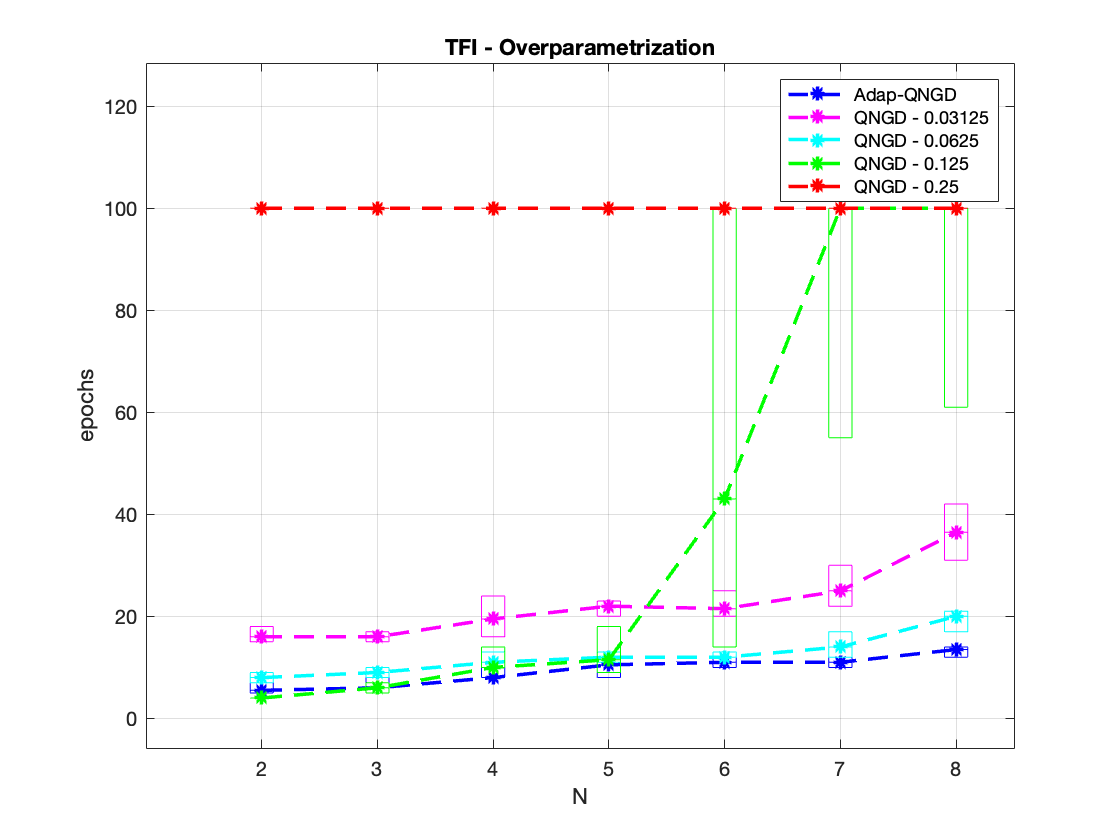}
    \caption{Comparison of median number of epochs needed of the ADAPT QNGD scheme with the QNGDS for varying rates, using Boxplots. The box around the median points indicates the containment of 25th to 75th percentiles.}
    \label{fig:overParam}
\end{figure} 
 
We conclude this section with another interesting result.We have simulated the TFI model in the context of  over-parametrization. Figure \ref{fig:overParam} shows the results obtained. We observe here a concerning aspect of the {\sc QNGD} scheme. Note that, the rate which performs superior to all other rates for lower $N$, eventually degrades in performance as $N$ is increased, when employing the {\sc QNGD} scheme. For instance, the rate of $0.125$ performs the best, on average, relative to other rates, for $N$ less than $5$, but significantly deteriorates for any $N$ larger than $5$. As a result, a new rate dominates amongst the set of rates, in terms of performance, as $N$ changes. This makes it exhausting to compute the best rate for every problem size (i.e., $N$) individually. {\sc AdaptQNGD}  was able to choose the most favourable rate for every problem size and performs equivalent to the best {\sc QNGD} performance without being aware of the optimal rate.

\section{Conclusion} \label{sec:conclusion}
In this paper, we have proposed an adaptive and efficient training scheme for variational quantum algorithms (such as QAOA and VQE) based on the Fubini-Study metric, in conjunction with Armijo line search rule.  Our results show that the new adaptive scheme outperforms {\sc QNGD} and mitigates its dependence  on the optimal step size. Our scheme captures the optimal dynamical change in the step size as the training proceeds. In addition, our results also show that 
 a simple gradient descent scheme (with respect to the Euclidean metric) equipped with the adaptive scheme, can yield performances similar to the {\sc QNGD} with optimal step size.  
 
More importantly, the results presented provide further confirmation of the importance of differential geometry in variational quantum computations. This is not surprising given the fact that this type of computation is  optimization of smooth functions on Riemannian manifolds. More generally, we foresee that  advanced mathematics will feature centrally in  the NISQ era in guiding the design of faster and more efficient practical algorithms.  

\section{Acknowledgement}
 We gratefully acknowledge the very helpful discussions with Nick Chancellor, and the rest of the Advanced Technology Team at QCI.  
\bibliographystyle{unsrt}
\bibliography{references}

\section{Appendix: Connection With $A-$stability Theory \cite{dahlquist}}
 
Let $f(\boldTheta) = \langle \varphi(\boldTheta)|H|\varphi(\boldTheta) \rangle$ be the expectation value where $|\varphi(\boldTheta)\rangle = U(\boldTheta) |0\rangle$ and $H$ is a Hamiltonian of interest (for instance, $U(\theta)$ is the QAOA ansatz and $H$ is a classical Hamiltonian i.e., a QUBO).  Let us also assume that $\boldTheta^*$ is a local minimum for the expectation value function $f(\boldTheta)$. One can  think of the {\it dynamical update} of the ansatz parameters  as the flow of the following dynamical system
\begin{system}\label{flow}
\Frac{d\boldTheta(\eps)}{d\eps} &=& - \nabla f (\boldTheta(\eps)),\\[3mm]
\boldTheta(\eps) &=& \boldTheta_0.
\end{system}%
 In the vicinity of $\boldTheta^*$, where $\boldTheta_0$ is assumed to belong,  we linearize the gradient flow above into the following linear ODE system (Dahlquist equations)
 \begin{system}\label{linflow}
\Frac{d\boldTheta(\eps)}{d\eps} &=& - H_f(\boldTheta^*) \cdot (\boldTheta-\boldTheta^*), \\[3mm]
\boldTheta(\eps) &=& \boldTheta_0,
\end{system}%
with $H_f(\boldTheta^*)$ the Hessian of $f$, which is positive semi-definite because $\boldTheta^*$ is a local minimum.  If we apply the forward Euler method, we get
 \begin{equation}
     \boldTheta_{k+1} = (Id -h H_f(\boldTheta_0) )\boldTheta_k =  (Id -h H_f(\boldTheta_0) )^k\boldTheta_0.
 \end{equation}
The term $R = (Id -h H_f(\boldTheta_0) )$ is the {\it stability function} of the forward Euler method and is required to satisfy $||R||<1$  for the method to converge.  Staying in   the vicinity of a local minimum, the optimal rate for the forward Euler method is 
 $h = {\sum \lambda_i}/{\sum \lambda_i^2 },$  where   $\lambda_i$ are the eigenvalues of the Hessian matrix $H_f(\boldTheta_0).$ This comes from the fact that   
the eigenvalues of the stability function $R$ are exactly the real numbers $1-h \lambda_i$, and subsequently, the optimal rate can be obtained with 
\begin{equation}
    argmin_{h>0} ||R|| = argmin_{h>0}  \sum_i (1-h \lambda_i)^2.
\end{equation}
Since Fubini-Study tensor is a positive semi-definite, after  the change of variables $\tau = F^{-1}(\boldTheta_0) \boldTheta$, the formula  above can also be applied to the new Hessian $H_f(\tau) = {F^{-1}}^t H_f F^{-1}$. 
Notice this change of coordinates changes the Riemannian manifold from the Euclidean space to the Fubini-Study manifold, and the dynamical update (the gradient flow (\ref{flow})) changes accordingly.

\begin{figure}[h]
    \centering
    \includegraphics[scale=0.20]{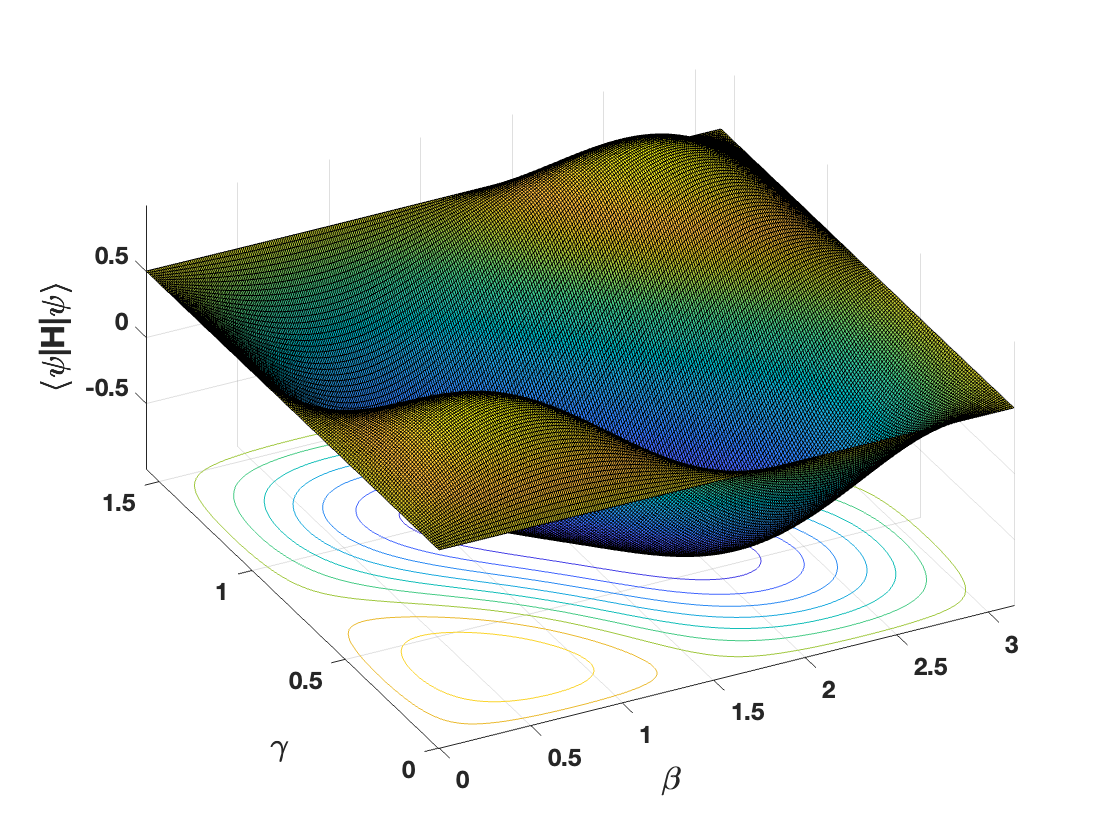}  
    \caption{Expectation value landscape for the simple example of QAOA depth 1 solving the max cut problem for a triangle. The level sets around the two symmetric global minima are not exactly circular, which makes the gradient descent to slightly slowdown.  }
    \label{triangle}
\end{figure}

Now consider depth one QAOA solving the max-cut problem for a triangle 
 \begin{eqnarray}\nonumber
   U(\boldTheta)=U(\beta, \gamma) &=& e^{-i \Frac{\beta}{2} (X_1 + X_2 + X_3)}\\\nonumber
    && \times e^{-i \Frac{\gamma}{2} (1-Z_1Z_2 - Z_1Z_3 - Z_2Z_3)},   \nonumber
 \end{eqnarray}
 with $(\beta, \gamma)\in [0, \pi/2]\times [0, \pi]$. 
Starting from the uniform superposition, we obtain the expectation value depicted in Figure \ref{triangle}, and the Fubini-Study tensor:
 \begin{equation}
     F(\boldTheta) = \left[\begin{array}{cc}
2-2 \cos \! \left(2 \boldTheta_{2}\right) & 0 
\\
 0 & \frac{1}{4} 
\end{array}\right]
 \end{equation}
A direct application of the optimal rate formula above gives $h = 0.093$ with $R=0.602$ when Fubini-Study is used,  and a slower rate $h = 0.022$ and $R=0.762$. In other words, Fubini-Study metric captures the correct geometry of the parameter space and makes the  gradient descent converge faster. 


\end{document}